\documentclass[openacc]{rstransa}

\usepackage{times}
\usepackage{color} 
\usepackage{booktabs}
\usepackage{lineno}
\usepackage{url}
\usepackage{pdfpages}
\usepackage{setspace} 
\usepackage{lineno}
\usepackage{subcaption}
\usepackage{amsmath}
\usepackage{amssymb}
\usepackage[comma,sort&compress]{natbib}

\titlehead{Research}

\begin{document}

\title{Evolution of fairness in hybrid populations with specialised AI agents}

\author{
Zhao Song$^{1}$, Theodor Cimpeanu$^{2}$, Chen Shen$^{3}$ and The Anh Han$^{1}$}

\address{$^{1}$Teesside University, United Kingdom.\\
$^{2}$University of Stirling, United Kingdom\\
$^{3}$Faculty of Engineering Sciences, Kyushu University, Japan}

\subject{Evolution, Prosociality, Human-AI interaction}


\corres{The Anh Han\\
\email{t.han@tees.ac.uk}}

\begin{abstract}
Fairness in hybrid societies hinges on a simple choice: should AI be a generous host or a strict gatekeeper? Moving beyond symmetric models, we show that asymmetric social structures--like those in hiring, regulation, and negotiation--AI that guards fairness outperforms AI that gifts it. We bridge this gap with a bipartite hybrid population model of the Ultimatum Game, separating humans and AI into distinct proposer and receiver groups. We first introduce Samaritan AI agents, which act as either unconditional fair proposers or strict receivers. Our results reveal a striking asymmetry: Samaritan AI receivers drive population-wide fairness far more effectively than Samaritan AI proposers. To overcome the limitations of the Samaritan AI proposer, we design the Discriminatory AI proposer, which predicts co-players' expectations and only offers fair portions to those with high acceptance thresholds. Our results demonstrate that this Discriminatory AI outperforms both types of Samaritan AI, especially in strong selection scenarios. It not only sustains fairness across both populations but also significantly lowers the critical mass of agents required to reach an equitable steady state.  By transitioning from unconditional modelling to strategic enforcement, our work provides a pivotal framework for deploying asymmetric AIs in the increasingly hybrid society.

\end{abstract}


\begin{fmtext}


\end{fmtext}


\maketitle
\section{Introduction}

As artificial intelligence (AI) transitions from a passive tool to an active participant in social and economic systems, a critical question emerges: how does its integration reshape foundational human behaviours such as fairness and cooperation~\citep{ke2022bad,alawadhi2021emergence,march2021human}? While fairness is a well-documented driver of human decision-making, the impact of AI on prosocial norms in hybrid human-AI societies remains poorly understood, particularly in contexts where social roles are asymmetric. 

Classical economics postulates fully rational, self-interested actors, but human social interactions pervasively deviate towards pro-sociality. The Ultimatum Game (UG) starkly captures this tension: a proposer offers a split of a resource, and a receiver decides whether to accept or reject it~\citep{guth1982experimental}. Although rational choice predicts minimal offers and universal acceptance, humans frequently reject unfair splits and make generous offers to avoid rejection--a pattern attributed to fairness preferences and inequity aversion~\citep{wallace2007heritability,solnick2001gender}.  This game has served as a canonical paradigm for studying the evolution of social norms.

Today, AI agents are increasingly embedded in roles traditionally occupied by humans--as hiring algorithms, negotiation bots, or automated regulators. This shift requires moving beyond purely human-centric studies to examine hybrid populations where humans interact with strategically programmed AI \citep{mu2024multi,paiva2018engineering,kobis2025delegation}. However, existing research largely assumes role symmetry, where individuals occupy both positions of power in interactions. This overlooks a fundamental reality: many real-world interactions are characterised by persistent, specialised role asymmetries (e.g., employer-employee, service provider-client) \citep{dong2024toward,cimpeanu2021cost}. In such bipartite structures, achieving collective fairness becomes more challenging, often requiring targeted intervention~\citep{cimpeanu2021cost,teixeira2021eliciting,cimpeanu2023social}. Furthermore, human behaviour toward machines often differs from toward humans—a `machine penalty' may reduce cooperation~\citep{bonnefon2024moral}. While human-like AI can mitigate this and even simple AI `zealots' can promote prosocial norms \citep{zimmaro2024emergence,sharma2023small,santos2019evolution}, our understanding of AI's impact remains incomplete in \textit{role-asymmetric} settings, where AI is locked into specialised positions.

It remains an open question how role-specialised AI agents, locked into either proposer or receiver positions, influence the evolution of fairness in hybrid populations. Does an AI that models fairness (as a generous proposer) differ in efficacy from one that enforces it (as a strict receiver)? Can they work synergistically? To address this, we develop a bipartite hybrid population model of the Ultimatum Game \citep{nowak2000fairness}, where humans and AI agents are explicitly partitioned into separate proposer and receiver populations. We first evaluate Samaritan AI agents, which are programmed with fixed, unconditional fair strategies. By restricting these agents to functioning solely as either norm-modellers (proposers) or norm-enforcers (receivers), we examine their respective capacities to shape social norms.

Our results reveal a fundamental asymmetry: Samaritan AI receivers are significantly more potent at driving population-wide fairness than Samaritan AI proposers. Through theoretical analysis in the finite well-mixed population \citep{nowak2004emergence,key:imhof2005}, we find that the AI receiver induces full fairness across both populations, whereas AI proposers fail to incentivise fairness of human receivers, particularly collapsing in strong selection scenarios. To address this limitation, we introduce a Discriminatory AI proposer that moves beyond unconditional generosity by predicting the receiver’s expectation, offering the minimal acceptable amount. 
 For this analysis, AI agents are equipped  with advanced decision-making capabilities: they can infer the behaviour or intentions of human receivers they interact with, for instance by learning from past actions \citep{han2013state,sukthankar2014plan}, recognising emotions \citep{tarnowski2017emotion}, predicting personality traits \citep{wright2026assessing}, or using reputation scores \citep{kas2022role}. Such capacities have been demonstrated with high accuracy using state-of-the-art AI methods, including deep learning and LLMs \citep{chowdary2023deep}. Recent advances in large language models, such as ChatGPT, also show increasing theory-of-mind–like abilities and inference of others’ mental states \citep{strachan2024testing}. In our model, we idealise these abilities by assuming AI proposers know receivers’ intended actions in advance, thereby removing response uncertainty and yielding a tractable yet meaningful simplification.

We find that this adaptive approach outperforms both Samaritan types across a wide range of parameter settings; it not only secures fairness across populations but also requires a substantially smaller critical mass of AI agents to sustain equitable outcomes, especially in strong selection scenarios. These findings offer a pivotal design principle for hybrid human-AI societies: to maximise impact and cost-effectiveness, AI interventions should prioritise strategic enforcement over unconditional modelling. This work provides a framework for understanding and designing AI’s role in the complex, asymmetric social structures that define our increasingly hybrid societies.


\section{Models and Methods}
\begin{figure}
    \centering
    \includegraphics[width=\linewidth]{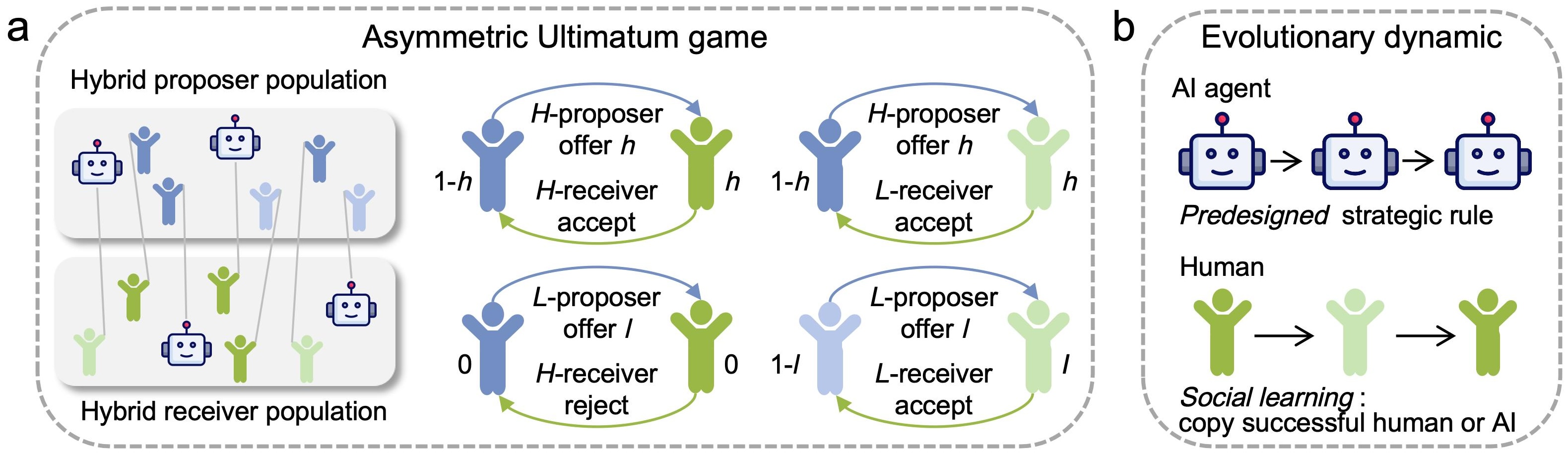}
    \caption{
    \textbf{Schematic of the Ultimatum Game in hybrid human-AI populations.}
    The model consists of two populations: the proposer population, where players choose between a high ($h$) and a low ($l$) offer, the and receiver population, where players have either a high ($h$) or low ($l$) acceptance threshold. 
    Each population can be a hybrid population consisting of human players and AI agents. 
    AI agents are pre-designed with a strategic rule. 
    Human players update their strategy following social learning to learn the successful strategy within their population.
    }
    \label{fig: model}
\end{figure}

\subsection{Ultimatum Game (UG)}

We model player interactions using the one-shot Ultimatum Game (UG) \citep{nowak2000fairness,page2000spatial}, Figure \ref{fig: model}a. In this two-player game, a sum of money, normalised to 1, is divided among players. The proposer offers a portion $p\in[0,1]$ to the receiver. The receiver, whose minimum acceptance threshold is $q\in[0,1]$, can either accept or reject the offer. If the offer is accepted ($p\geq q$), then the receiver gets $p$ and the proposer gets the rest $1-p$. If the offer is rejected, both players receive zero. In our model, we consider a discrete strategy space. Proposers can make either a low offer ($L_P$), $p=l$, or a high (fair) offer ($H_P$), $p=h$. Receivers can have either a low threshold ($L_R$), $q = l$, or a high threshold ($H_R$), $q = h$. We assume $0<l<h$. Following empirical evidence~\citep{guth1982experimental,rand2013evolution,zisis2015generosity}, we set $h=0.5$ and $l=0.1$ unless specified. The resulting payoff matrix for the row player (receiver) and column player (proposer) is:
\begin{equation*}
\centering
\def\arraystretch{1.25}
\begin{tabular}{c|c c}
\toprule
& $H_R$ & $L_R$  \\
\hline
$H_P$ & $1-h, h$ & $1-h, h$ \\
$L_P$ &  $0,0$ & $1-l, l$  \\
\bottomrule
\end{tabular}
\end{equation*}

\subsection{Hybrid human-AI populations: The Samaritan AI}
We consider the well-mixed, finite bipartite population, comprising a proposer population and a receiver population, Figure \ref{fig: model}b. Interactions occur globally between these two populations. Each population is a hybrid population consisting of human players and AI agents. Let $N_P$ and $N_R$ be the numbers of human proposers and receivers, respectively. The number of $H$-proposers and $H$-receivers in human players is denoted by $k_P$ and $k_R$. Additionally, there are  $M_P$ and $M_R$ AI proposers and receivers. 
In our baseline model, we introduce the Samaritan AI.
There agents adopt pre-designed, fixed prosocial behaviour: i) Samaritan AI proposers always adopt $H_P$, offering $p=h$; ii) Samaritan AI receivers always adopt $H_R$, setting the acceptance threshold of $q=h$.

\subsection{Evolutionary dynamics in finite populations}
Human players update their strategies through social learning, where players' payoffs represent their \textit{social success} and they tend to imitate the successful strategy \citep{sigmund2010calculus,hofbauer1998evolutionary}. Depending on the composition of the opposing population, the average payoffs for $H_P$ and $L_P$ are:
\begin{equation}
    \begin{split}
        &\pi_{H_P} = 1-h , \\
        &\pi_{L_P} = \frac{(1-l)  (N_R - k_R )}{N_R + M_R}.
    \end{split}
\end{equation}
The average payoffs for $H_R$ and $L_R$ are:
\begin{equation}
    \begin{split}
        &\pi_{H_R} = \frac{h  (k_P + M_P)}{N_P + M_P} ,\\
        &\pi_{L_R} = \frac{h(k_P+M_P)+l(N_P-k_P)}{N_P+M_P}.
    \end{split}
\end{equation}
While AI agents do not alter their strategies over time, the evolutionary dynamics of a human player's strategy follow a Moran process with pairwise comparison. At each time step, a human player $A$ with payoff $\pi_{A}$ is randomly selected to update the strategy. A role model, player $B$ with payoff $\pi_{B}$ is then chosen from the entire corresponding population (including humans and AIs). Player $A$ adopts player $B$'s strategy with a probability $p_{A,B}$ given by the Fermi function \citep{traulsen2006}:
\begin{equation}
    f_{A,B} = (1+e^{-\beta(\pi_{B}-\pi_{A})})^{-1},
\end{equation}
where $\beta$ is the intensity of selection that accounts for how much the players base on the payoff difference between the role model and themselves in social learning. For $\beta \rightarrow 0$, imitation is random; For large $\beta$, players almost always copy others with a higher payoff.

In the absence of mutations or exploration, the end states of evolution are inevitably monomorphic: once such a state is reached, it cannot be escaped through imitation. We thus further assume that with a certain mutation probability,  an agent switches randomly to a different strategy without imitating another agent.  In the limit of small mutation rates, the dynamics will proceed with, at most, two strategies in the population, such that the behavioural dynamics can be conveniently described by a Markov chain, where each state represents a monomorphic population, whereas the transition probabilities are given by the fixation probability of a single mutant \citep{imhof2005evolutionary,nowak2004emergence}. The resulting Markov chain has a stationary distribution, which characterises the average time the population spends in each of these monomorphic end states.

Now, the transition probability of the number of (human) $H$-proposer increasing ($T^+$) or decreasing ($T^-$) by one can be defined as:
\begin{equation}
    \begin{split}
    T^{+}(k_P) &= \frac{N_P-k_P}{N_P}\frac{k_P+ M_P}{N_P+M_P} f_{L_P,H_P},\\
    T^{-}(k_P) &= \frac{k_P}{N_P}\frac{N_P-k_P }{N_P+M_P} f_{H_P,L_P}.
\end{split}
\end{equation}
Here, $T^{+}(k_P)$ is the probability that one of the ($N_P-k_P$) L-proposers is selected to update its strategy (first term), chooses one of the ($k_P+M_P$) $H$-proposers as a role model from the population (second term), and adopts the strategy (third term). A similar logic applies to $T^{-}(k_P)$.

Similarly, the transition probability of the number of $H$-receivers increasing ($T^+$) or decreasing ($T^-$) by one can be defined as:
\begin{equation}
    \begin{split}
    T^{+}(k_R) &= \frac{N_R-k_R}{N_R}\frac{k_R+ M_R}{N_R+M_R} f_{L_R,H_R},\\
    T^{-}(k_R) &= \frac{k_R}{N_R}\frac{N_R-k_R }{N_R+M_R} f_{H_R,L_R}.
\end{split}
\end{equation}

This allows us to calculate the fixation probability $\rho_{B,A}$ of a single mutant with a strategy $A$ in a resident population with strategy $B$ \citep{traulsen2006,key:novaknature2004}:
\begin{equation} 
\label{eq:fixprob} 
\rho_{B,A} = \left(1 + \sum_{i = 1}^{N-1} \prod_{j = 1}^i \frac{T^-(j)}{T^+(j)}\right)^{-1},
\end{equation} 
where $N$ is the number of human players set as $N_P=N_R=100$.
The fixation probability ($\rho_{ij}$) denotes the likelihood that a population transitions from state ($i$) to a different state ($j$) when a mutant in one of the populations adopts an alternative strategy ($s$). This fixation probability is divided by the number of populations (two), reflecting that interactions involve four players at a time \citep{encarnaccao2016paradigm,ALALAWI2026129627}.

The population dynamics can be in one of four homogeneous states: $HH$, $HL$, $LH$, and $LL$. 
Assuming a small mutation limit, where any mutant either fixates or goes extinct before another mutation occurs \citep{nowak2004emergence,imhof2005evolutionary},   the fixation probabilities $\rho_{ij}$ define the transition matrix, $A$, of a Markov process between the four homogeneous population states~\citep{fudenberg2006imitation,karlin2014first}. 
The transition probability from state $i$ to state $j$ is given by $\Lambda_{ij}$:
\begin{eqnarray}\label{eq:2.6}
\Lambda_{ij,j\neq i}=\frac{\rho_{ji}}{2}\hspace{1mm} \text{ and } \hspace{1mm} \Lambda_{ii}= 1- \sum_{j=1,j\neq i}^s \Lambda_{ij},
\end{eqnarray}
The long-term outcome of this evolutionary process is captured by the stationary distribution of this matrix. Mathematically, this distribution corresponds to the normalised eigenvector of the transposed transition matrix with an eigenvalue of 1. In our analysis, this stationary distribution represents the fraction of time the population is expected to spend in each monomorphic state in the long run. We therefore use it as our primary measure of a strategy's long-term evolutionary success.


\subsection{The Discriminatory AI}
Beyond the Samaritan AI proposer, which always adopts a $h$ strategy, we introduce Discriminatory AI proposers only offer $h$ to the receiver with an $h$ predicted threshold, otherwise $l$ offer. In addition, such Discriminatory AI is perceived by human proposers as a H-proposer with probability $\alpha = \frac{k_R + M_R}{N_R + M_R}$, where $k_R$ is the number of receivers adopting $h$. 

Under this framework, the average payoffs for receivers can be rewritten as follows:
\begin{equation}
    \begin{split}
        & \pi_{H_R}' = \frac{h(k_P+ M_P)}{N_P+M_P}, \\
        & \pi_{L_R}' = \frac{hk_P+l(N_P-k_P+M_P)}{N_P+M_P}.
    \end{split}
\end{equation}
Besides, the transition probabilities $T^{\pm}$ for human proposers and receivers are adjusted as:
\begin{equation}
\begin{split}
T^{+}(k_P) &= \frac{N_P-k_P}{N_P}\frac{k_P+ \alpha  M_P}{N_P+M_P} f_{L_P,H_P},\\
T^{-}(k_P) &= \frac{k_P}{N_P}\frac{N_P-k_P + (1-\alpha) M_P}{N_P+M_P} f_{H_P,L_P},
\end{split}
\end{equation}
\begin{equation}
    \begin{split}
    T^{+}(k_R) &= \frac{N_R-k_R}{N_R}\frac{k_R+ M_R}{N_R+M_R} f_{L_R,H_R}',\\
    T^{-}(k_R) &= \frac{k_R}{N_R}\frac{N_R-k_R }{N_R+M_R} f_{H_R,L_R}',
\end{split}
\end{equation}
where $f_{L_R,H_R}'=1/(1+e^{-\beta(\pi_{H_R}'-\pi_{L_R}')})$ and $f_{H_R,L_R}'=1/(1+e^{-\beta(\pi_{L_R}'-\pi_{H_R}')})$.

\section{Results}
\subsection{The Samaritan AI}
\begin{figure}[htp]
    \centering
    \includegraphics[width=1\linewidth]{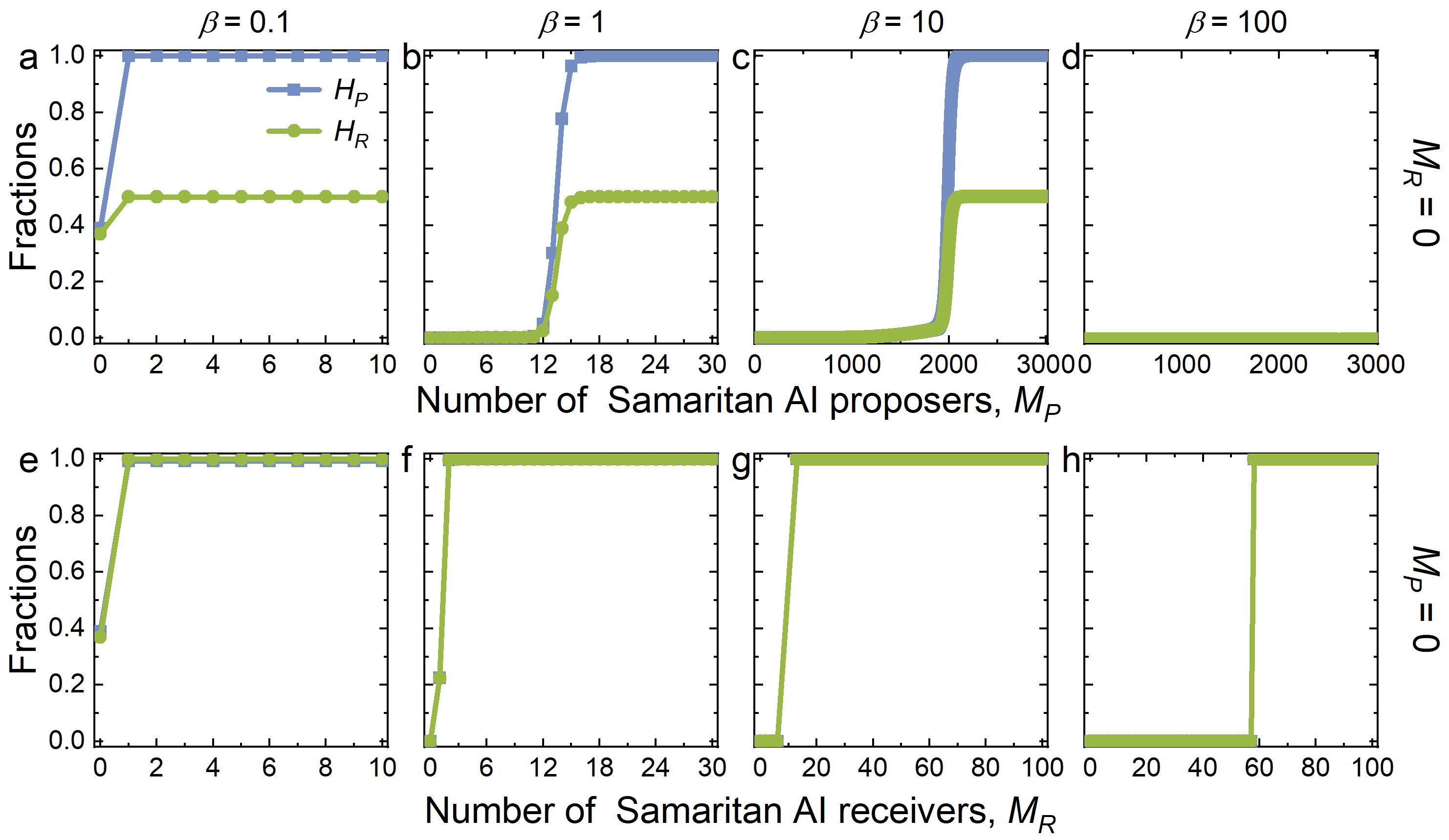}
    \caption{
    \textbf{
    AI receivers outperform AI proposers in enhancing fairness in both populations. }
    Panels in the top row show the fraction of $H$-proposers and $H$-receivers among human players as a function of the number of AI proposers with different selection intensities, respectively.
    Panels in the bottom row show the fraction of $H$-proposers and $H$-receivers as a function of the number of AI receivers with different selection intensities, respectively.
    Parameters are set as $h=0.5$, $l=0.1$, $\beta \in \{0.1, 1, 10, 100\}$ from left to the right column, respectively, $M_R=0$ in the top row and $M_P=0$ in the bottom row.}
    \label{fig1}
\end{figure}

\begin{figure}[htb]
    \centering
     \includegraphics[width=\linewidth]{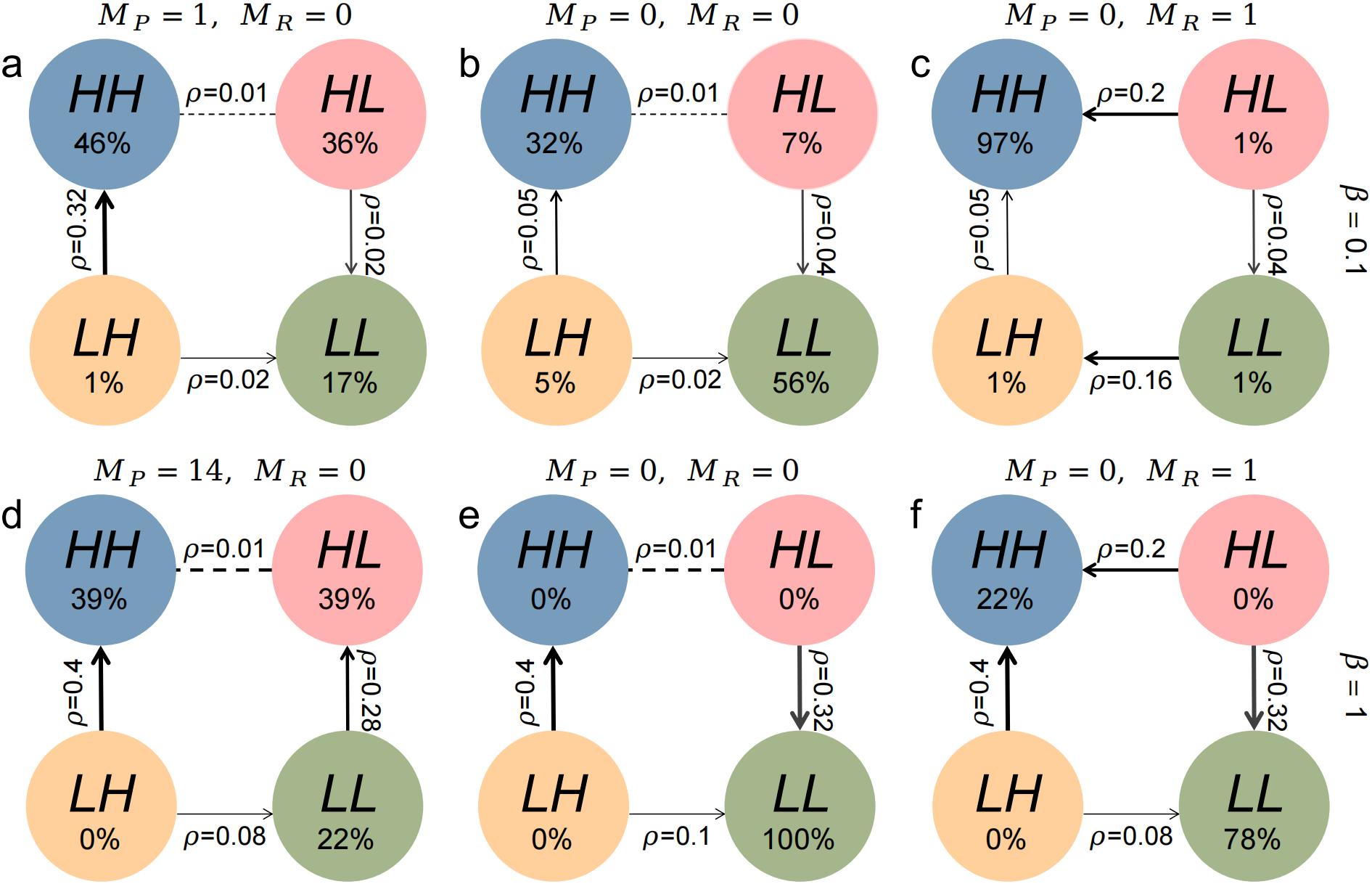}
    \caption{
    \textbf{AI receivers effectively alter the evolutionary direction to enhance fairness.}
    The percentages indicate the stationary distribution of each state. The arrows represent the transition direction between paired strategies, highlighting the transition probabilities and the stronger direction of transitions. Dashed lines represent neutral transitions. Horizontal (vertical) lines represent transition within the receiver (proposer) population. 
    Parameters are set as $h=0.5$, $l=0.1$,  (a) $M_P=M_R=0$, $\beta=0.1$, (b) $M_P=1$, $M_R=0$, $\beta=0.1$, (c) $M_P=0$, $M_R=1$, $\beta=0.1$, (d) $M_P=M_R=0$, $\beta=1$, (e) $M_P=14, M_R=0$, $\beta=1$, and (f) $M_P=0, M_R=1$, $\beta=1$.}
    \label{fig2}
\end{figure}

\paragraph{AI receivers are more effective in enhancing fairness.} Introducing AI receivers is a more effective approach for promoting fairness than introducing AI proposers across selection intensities. The presence of AI receivers consistently drives human players in both populations toward full fairness, while AI proposers at best encourage full fairness among human proposers only, as shown in Figure \ref{fig1}. Under weak selection intensity ($\beta=0.1$, first column in Figure \ref{fig1}), for example, a single AI receiver induces full fairness in both human populations, whereas a single AI proposer achieves full fairness only among proposers and partial fairness among receivers. As selection intensity increases, more AI receivers are required to ensure full fairness; at $\beta=100$ (last column in Figure \ref{fig1}), a group of around $M_R \geq 58$ AI receivers is needed. In contrast, AI proposers become completely ineffective under strong selection, failing to promote fairness in either population.

The analysis of the underlying transition dynamics reveals the effectiveness of AI receivers: they fundamentally alter the evolutionary pathways to favour mutual fairness. Under weak selection ($\beta=0.1$), the unfair $LL$ state is heavily favoured without AI agents (Figure \ref{fig2}a). A single AI proposer strengthens the transition from $LH$ to $HH$ ($\rho=0.32$), but the system still settles in the partially fair $HL$ state (Figure \ref{fig2}b). In contrast, a single AI receiver creates powerful new pathways to the fully fair $HH$ state-both from $HL$ ($\rho=0.2$) and by reversing the flow between $LL$ and $LH$ ($\rho=0.16$). 
These rewire the entire dynamic, making the $HH$ state the  dominant outcome (Figure \ref{fig2}c). This disparity is even more pronounced under strong selection, where the baseline population is completely trapped in the unfair $LL$ state (Figure \ref{fig2}d). Here, even a large group of AI proposers (around $M_P=14$) only creates a pathway to the partially fair $HL$ state. However, just one AI receiver is sufficient to create a direct escape route toward mutual fairness by enabling the critical transition from $HL$ to $HH$ (Figure \ref{fig2}e,f).
It is noteworthy that the transition probabilities between $HL$ and $HH$ do not depend on the value of $\beta$ since $\pi_{H_R} = \pi_{L_R} = h$ (and thus $f_{L_R,H_R} = f_{H_R,L_R} = 0.5$) when the proposer population adopts $H_R$. 

\begin{figure}[tb]
    \centering
        \includegraphics[width=\linewidth]{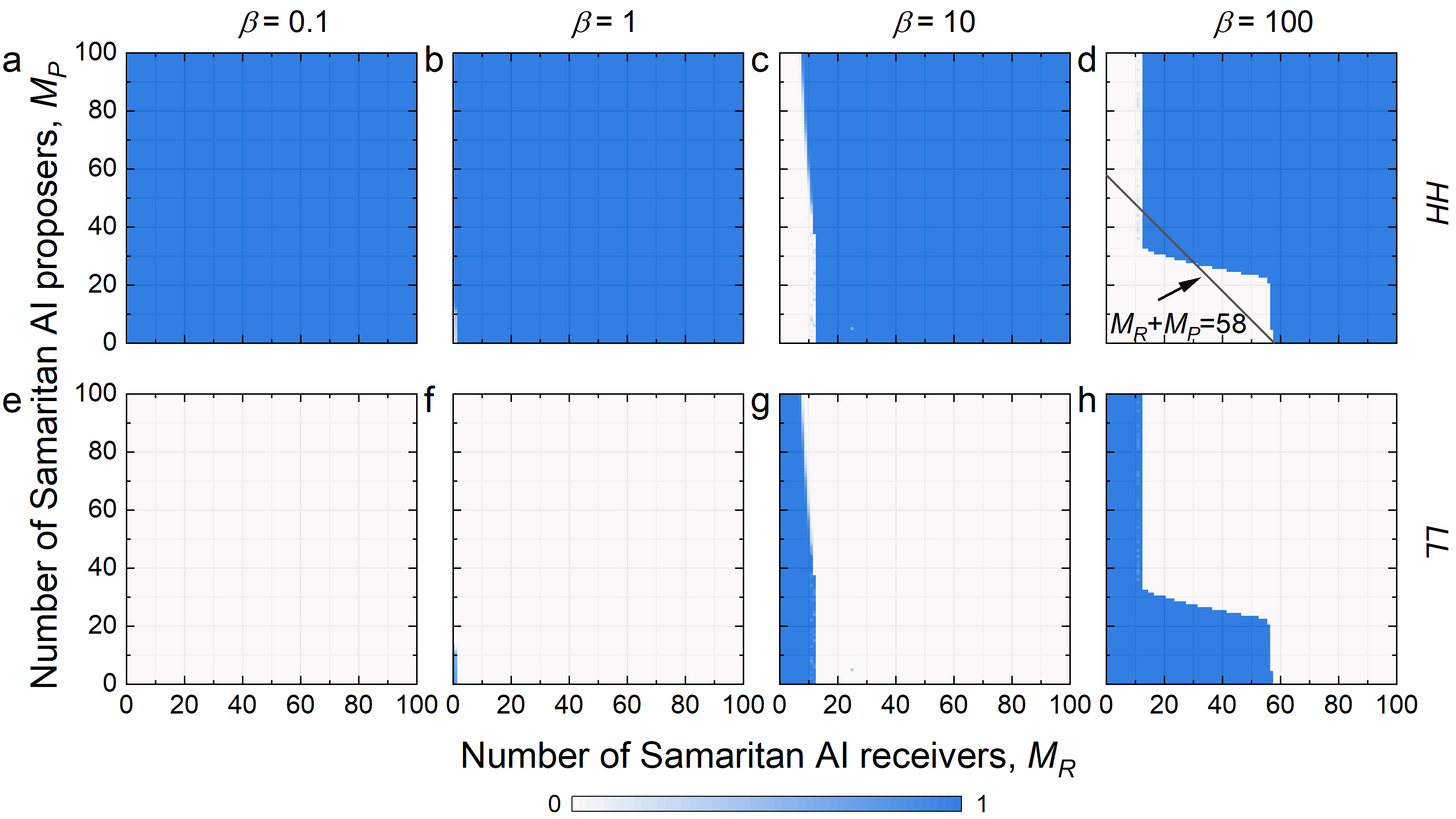}
    \caption{
    \textbf{The interplay of AI proposer and receiver shows better effectiveness at large selection intensity.}
    Panels in the top (bottom) row show the stationary distributions of $HH$ ($LL$) as a function of the numbers of AI receivers and AI proposers with different selection intensities, respectively. 
    The solid line in panel (m) represents the sum of AI agents, $M_R+M_P=58$.
    Parameters are set as $h=0.5$, $l=0.1$, $\beta=0.1$ in the first column, $\beta=1$ in the second column, $\beta=10$ in the third column, and $\beta=100$ in the last column.}
    \label{fig3}
\end{figure}

\paragraph{Interplay of AI proposers and AI receivers reduces the needed AIs.} While AI receivers outperform AI proposers when examined individually, it's worth noticing that a substantial number of AI receivers is required under high selection intensity. Given the potential costs of integrating AI agents into human populations, it becomes essential to explore more cost-effective methods for utilising AI. 

The combined introduction of AI proposers and receivers reveals a synergistic effect that reduces the total number of agents required to establish population-wide fairness, especially under strong selection. This effect is most apparent at high selection intensity ($\beta=100$). In this scenario, AI proposers alone are ineffective, while achieving fairness requires at least 58 AI receivers. When both types are introduced together, however, the boundary between the unfair  $LL$ and fair $HH$ states shows a distinct linear trade-off, following the relationship $M_P+M_R=58$ (the last column in Figure \ref{fig3}). This demonstrates that the total number of AI agents becomes the determining factor, allowing a mix of agents (e.g., 20 AI receivers and 38 AI proposers) to achieve the same outcome as a larger, single-type group. A similar synergy is observable at moderate selection ($\beta=10$), where the number of AI receivers required to ensure fairness is substantially reduced by the presence of AI proposers (the third column in Figure \ref{fig3}). Under weak selection ($\beta\leq 1$), the population achieves fairness regardless, with any small number of AI agents being sufficient to secure the $HH$ state (the first and second columns in Figure \ref{fig3}).

\begin{figure}[tb]
    \centering
        \includegraphics[width=1\linewidth]{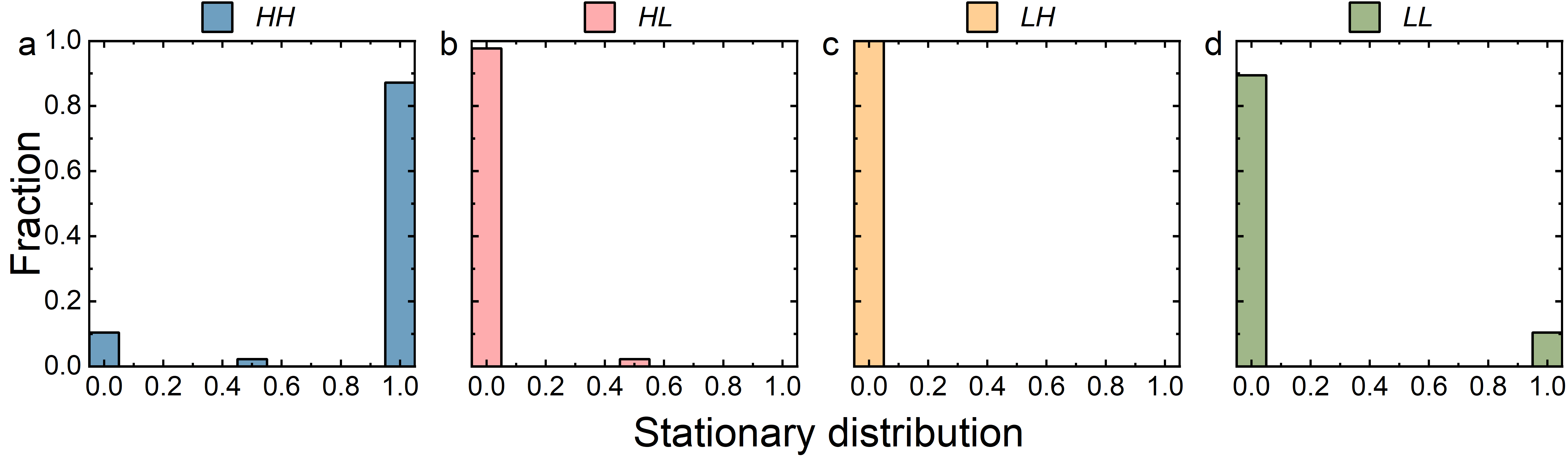}
    \caption{
    \textbf{AI agents robustly sustain mutual fairness across a wide range of game payoffs and selection intensities.}
    Shown are the frequencies of each stationary distribution.
    Numerical calculations results are obtained using $M_P \in [0,100]$ with step 5, $M_R \in [0,100]$ with step 5, $h \in[0.4,0.6]$ with step 0.01, $l \in[0.1,0.3]$ with step 0.01, and $\beta \in \{0.1, 1, 10, 100\}$.}
    \label{fig4}
\end{figure}
\paragraph{Robustness across parameters.} The results presented so far rely on a fixed set of game payoffs, where the high offer $h=0.5$ and the low offer $l=0.1$. To verify the generality of our conclusions, we now test the robustness of AI agents in enhancing fairness by exploring a wide range of game payoffs and selection intensities.

Our analysis confirms that the ability of AI agents to enhance fairness is robust across a wide range of game payoffs ($h,l$) and selection intensities ($\beta$). The system overwhelmingly converges to the mutually fair (HH) state for the vast majority of the tested parameter combinations (Figure \ref{fig4}a). Specifically, for approximately 88$\%$ of all scenarios explored, the stationary distribution of the $HH$ state is nearly 1.0. In contrast, the mutually unfair $LL$ is the dominant outcome in only about 10$\%$ of cases (Figure \ref{fig4}d). The partially fair states, $HL$ and $LH$, are rarely observed as stable long-term outcomes, indicating that the system consistently evolves towards one of the two extremes of full fairness or full unfairness.

\begin{figure}[tb]
    \centering
        \includegraphics[width=1\linewidth]{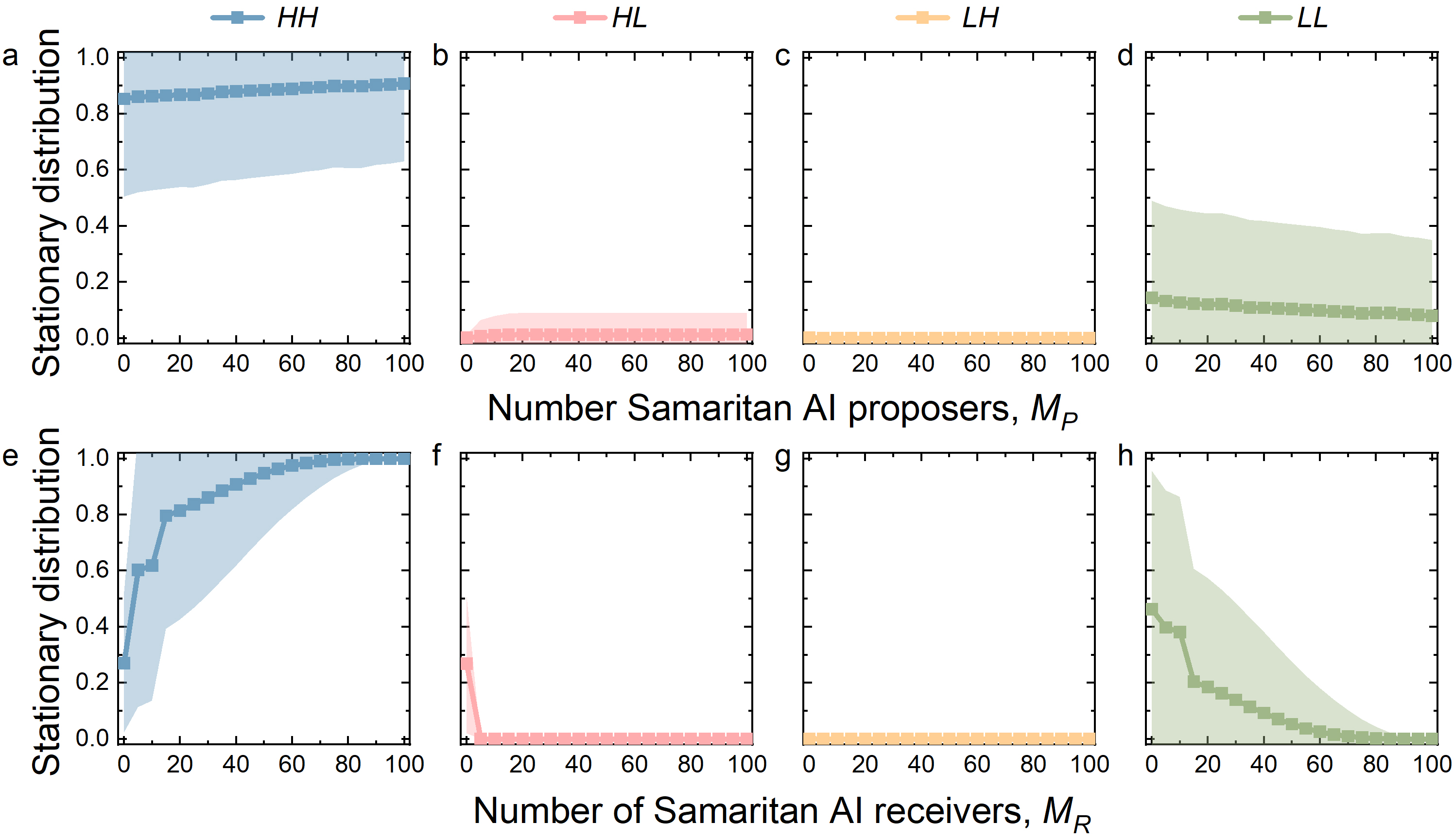}
    \caption{
    \textbf{While both AI types promote fairness, the effect of AI receivers is more pronounced across parameters.}
    Shown are the stationary distributions as a function of the number of FAIR AI across parameter sets. 
    Numerical calculations results are obtained using $M_P \in [0,100]$ with step 5, $M_R \in [0,100]$ with step 5, $h \in[0.4,0.6]$ with step 0.01, $l \in[0.1,0.3]$ with step 0.01, and $\beta \in \{0.1, 1, 10, 100\}$.}
    \label{fig5}
\end{figure}

This robustness analysis also reinforces the superior efficacy of AI receivers when results are averaged across the entire parameter space. The impact of AI receivers is particularly pronounced. As their number increases, the average stationary distribution of the fair $HH$ state climbs steeply from approximately 0.3 to 1 (Figure \ref{fig5}e). This trend is accompanied by a significant reduction in the variance of outcomes (the shaded region), indicating that AI receivers not only enhance fairness but also make it a more certain result across different conditions. Correspondingly, the unfair $LL$ state is driven towards complete elimination (Figure \ref{fig5}h). In contrast, the influence of AI proposers is more modest. Increasing their number results in a gradual increase in the average $HH$ state's prevalence, from roughly 0.85 to 0.95, as the unfair $LL$ state's prevalence remains largely unchanged around 0.1 (the top row in Figure \ref{fig5}).

\begin{figure}
    \centering
        \includegraphics[width=1\linewidth]{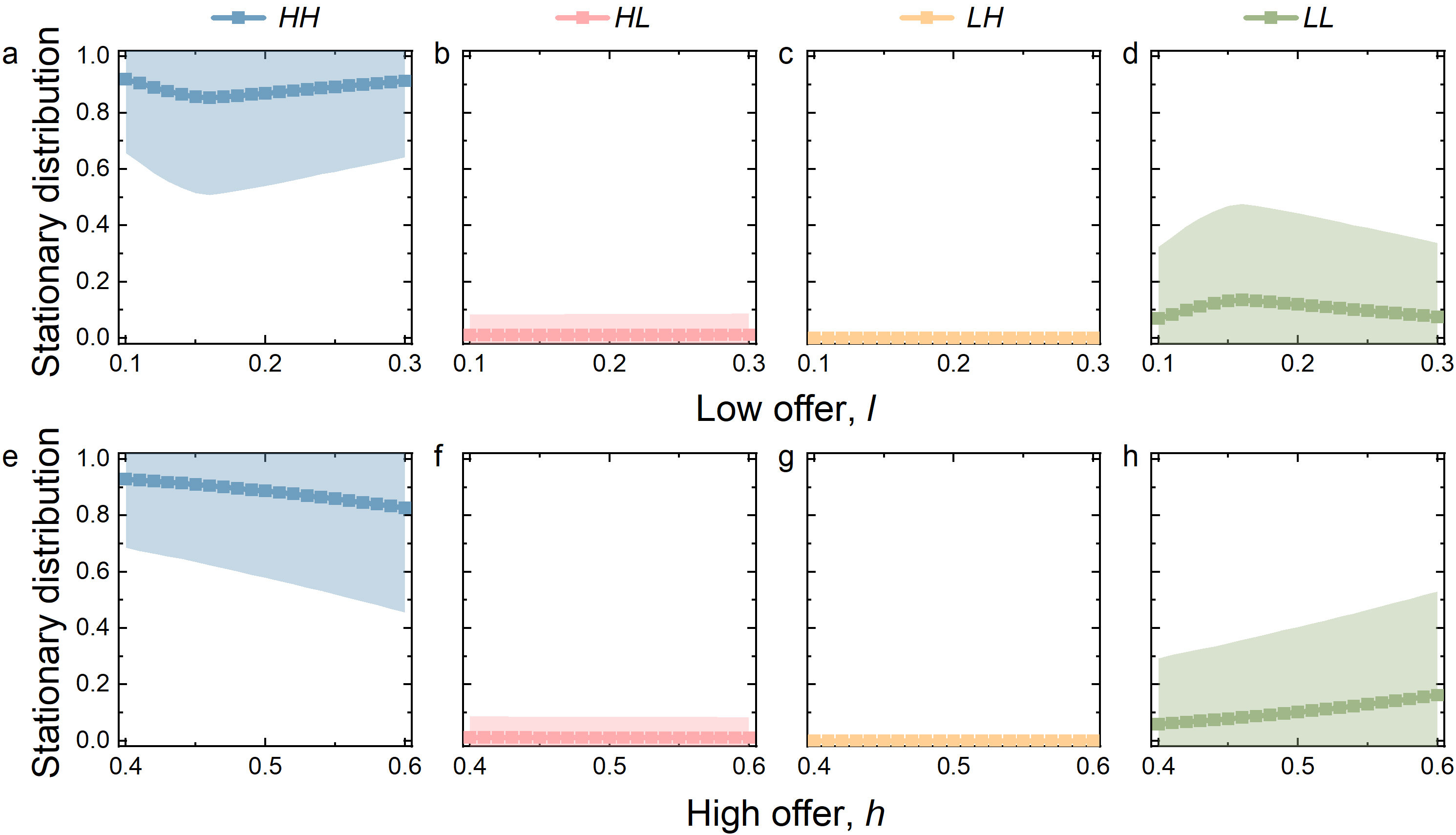}
    \caption{
    \textbf{Mutual fairness is robustly achieved when the low offer is minimal and the high offer is not excessively generous.}
    Shown are the stationary distributions as a function of high offer and low offer across parameter sets. 
    Numerical calculations results are obtained using $M_P \in [0,100]$ with step 5, $M_R \in [0,100]$ with step 5, $h \in[0.4,0.6]$ with step 0.01, $l \in[0.1,0.3]$ with step 0.01, and $\beta \in \{0.1, 1, 10, 100\}$.}
    \label{fig6}
\end{figure}

Finally, we find that mutual fairness is most robustly achieved when the low offer is minimal or the high offer is not excessively generous. Increasing the value of the low offer, $l$, which makes the 'unfair' strategy more rewarding for $L$-proposer, correlates with a lower average prevalence of the fair $HH$ state and a higher prevalence of the unfair $LL$ state (Figure \ref{fig6}a,d). A similar trend occurs for the high offer, $h$. As the fair offer becomes more generous (and thus more costly for proposers), the average occurrence of the $HH$ state also decreases, while the $LL$ state becomes more likely (Figure \ref{fig6}e,h). This suggests that the stability of the fully fair state is highest when the 'unfair' strategy is easily punishable (low $l$) and the 'fair' strategy is not overly costly for proposers (low $h$).

\subsection{The Discriminatory AI}
While  Samaritan AI agents' analysis  provides a baseline for the impact of simple AI on enhancing pro-sociality, the previous results highlight a critical limitation: Samaritan AI proposers fail to effectively incentivise human receivers to adopt high thresholds, especially under strong selection. To address this, we introduce the Discriminatory AI proposer.
In the following, we show how Discriminatory AI proposers influence the fairness.

\begin{figure}[tp]
    \centering
    \includegraphics[width=1\linewidth]{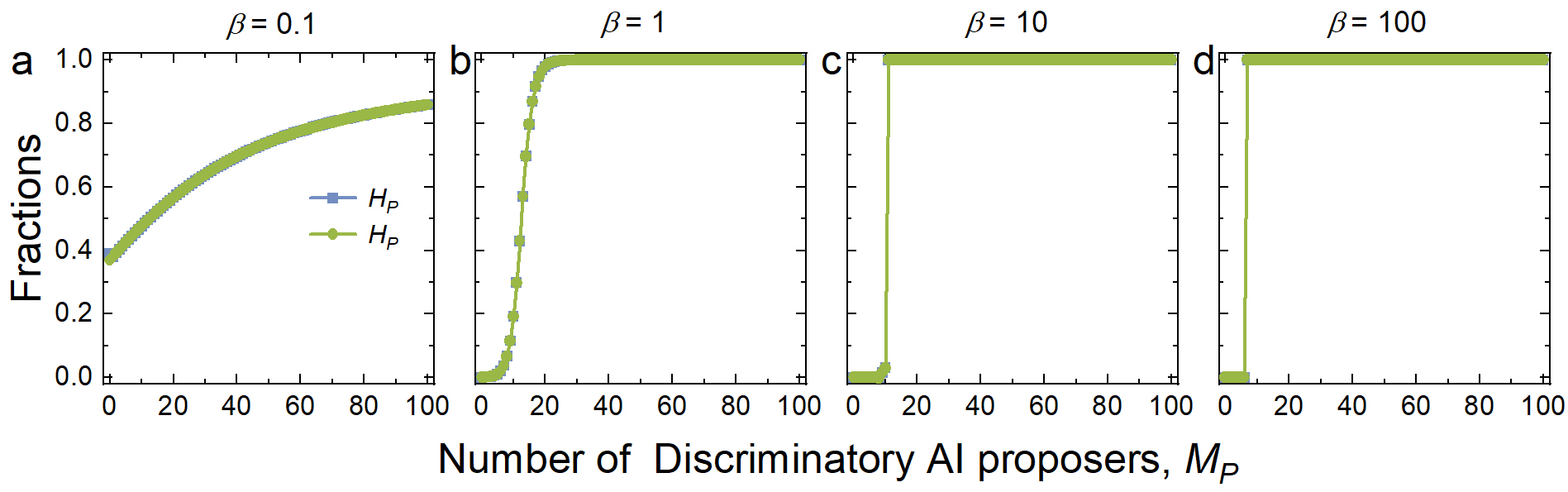}
    \caption{
    \textbf{
    Discriminatory AI proposers perform better in strong selection scenarios.}
    Panels show the fraction of $H$-proposers and $H$-receivers among human players as a function of the number of Discriminatory AI proposers with different selection intensities, respectively.
    Parameters are set as $h=0.5$, $l=0.1$, $\beta \in \{0.1, 1, 10, 100\}$ from left to the right column, respectively.}
    \label{figA1}
\end{figure}

\begin{figure}[htb]
    \centering
     \includegraphics[width=\linewidth]{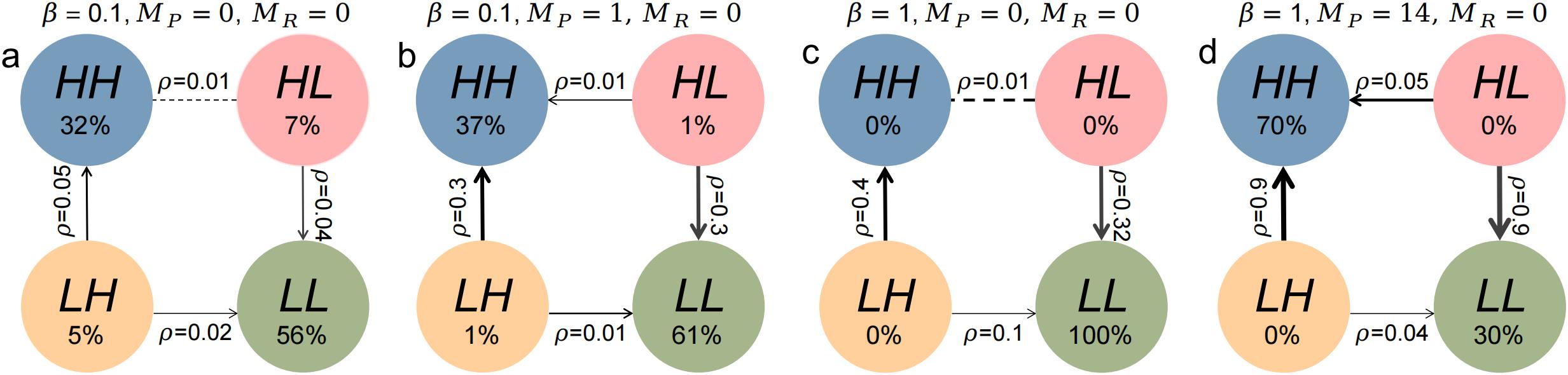}
    \caption{
    \textbf{Discriminatory AI proposers ensure the evolutionary direction towards HH.}
    The percentages indicate the stationary distribution of each state. The arrows represent the transition direction between paired strategies, highlighting the transition probabilities and the stronger direction of transitions. Dashed lines represent neutral transitions. Horizontal (vertical) lines represent transition within the receiver (proposer) population. 
    Parameters are set as $h=0.5$, $l=0.1$,  (a) $M_P=M_R=0$, $\beta=0.1$, (b) $M_P=1$, $M_R=0$, $\beta=0.1$, (c) $M_P=M_R=0$, $\beta=1$, and (d) $M_P=14, M_R=0$, $\beta=1$.}
    \label{figA2}
\end{figure}

\paragraph{Discriminatory AI proposers outperforms Samaritan AI for strong selection.}
Discriminatory AI proposers perform better in strong selection scenarios, outperforming both Samaritan types in establishing population-wide fairness with high efficiency. While Samaritan AI proposers at best encourage fairness only within the proposer population and become entirely ineffective under strong selection, Discriminatory AI proposers consistently drive both human populations toward full fairness. More importantly, compared to Samaritan AI receivers which require a substantial group of agents to maintain fairness in strong selection scenarios (e.g.,  $M_R \geq 58$ is required when $\beta=100$), as shown in Figure \ref{figA1}, Discriminatory AI proposers achieve the same outcome with significantly fewer agents. Specifically, when $\beta=100$, only a small group of approximately $M_P \geq 8$ Discriminatory AI proposers is sufficient to ensure the dominance of the fairness across populations. This highlights that by conditioning offers on receiver strategies, Discriminatory AI proposers not only overcome the ineffectiveness of their Samaritan counterparts but also offer a more cost-effective intervention than Samaritan receivers for fostering global pro-sociality under high selection pressure.

Discriminatory AI proposers fundamentally alter evolutionary pathways to favour mutual fairness, demonstrating a superior capacity to reshape population dynamics compared to Samaritan AI. Under weaker selection ($\beta=0.1$), while the absence of AI leads the system to favour the unfair $LL$ state (Figure \ref{figA2}a), the introduction of a single Discriminatory AI proposer strengthens the transition from $HL$ to $HH$ ($\rho=0.3$), resulting in the a higher likelihood of reaching the fully fair state compared to the baseline (Figure \ref{figA2}b). Crucially, unlike Samaritan AI proposers, which at best guided the population toward the partially fair $HL$ state ($\beta=1$), Discriminatory AI directly enables the transition to full $HH$ fairness. As shown in (Figure \ref{figA2}c and d), in a population otherwise doomed to 100\% unfair behaviour, a group of Discriminatory AI proposers reverses the evolutionary flow between $HL$ and $HH$ ($\rho=0.05$) and, more importantly, establishes a near-certain transition from $HL$ and $LH$ to $HH$ and $LL$ ($\rho=0.9$).  This suggests that by assessing and conditioning offers on receiver behaviour, Discriminatory AI proposers act as a more potent evolutionary catalyst than Samaritan types, securing the $HH$ state as the dominant outcome, specifically when selection pressure is strong.

\begin{figure}[tb]
    \centering
        \includegraphics[width=1\linewidth]{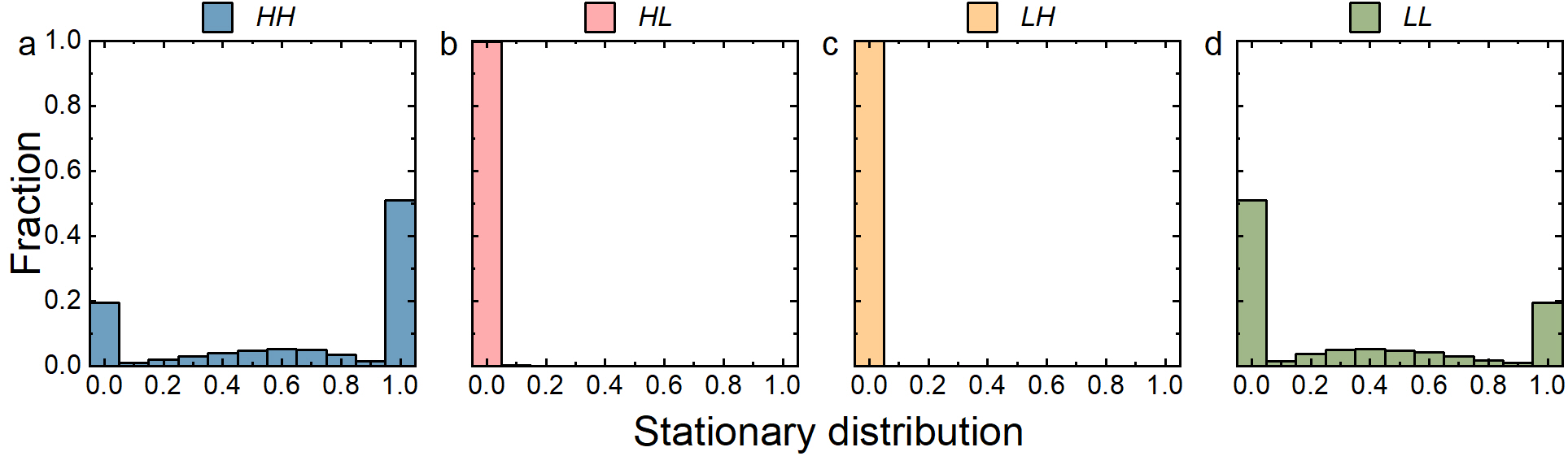}
    \caption{
    \textbf{Discriminatory AI proposers sustain mutual fairness across a wide range of parameters.}
    Shown are the frequencies of each stationary distribution.
    Numerical calculations results are obtained using $M_P \in [0,100]$ with step 5, $h \in[0.4,0.6]$ with step 0.01, $l \in[0.1,0.3]$ with step 0.01, and $\beta \in \{0.1, 1, 10, 100\}$.}
    \label{figA3}
\end{figure}

\begin{figure}[tb]
    \centering
        \includegraphics[width=1\linewidth]{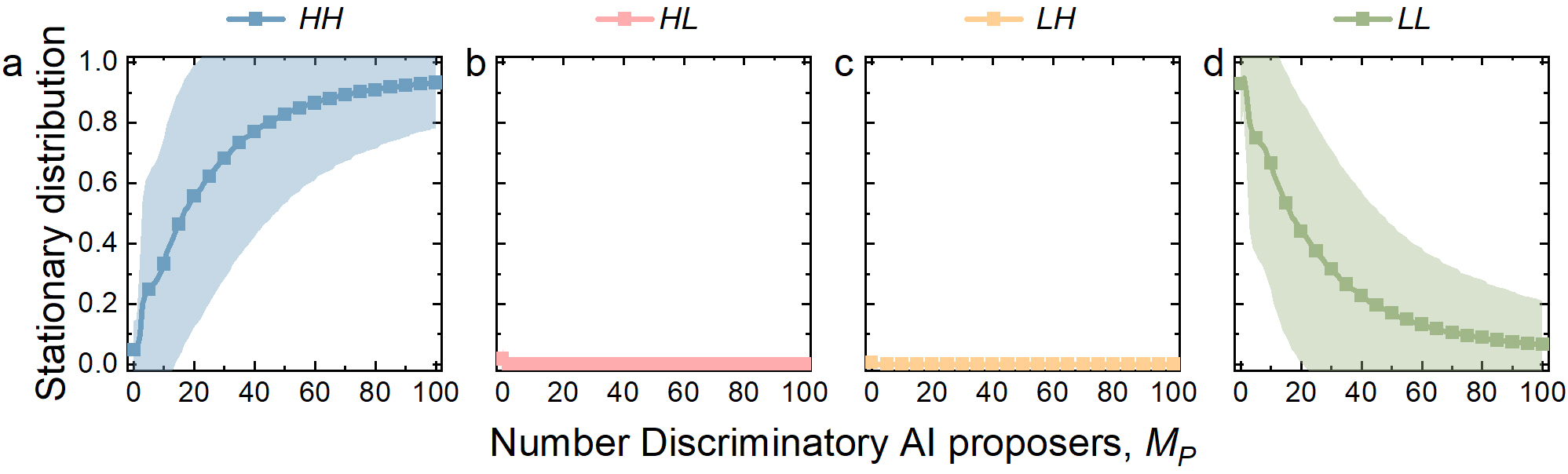}
    \caption{
    \textbf{The prevalence of mutual fairness increases monotonically with the number of Discriminatory AI proposers.}
    Shown are the stationary distributions as a function of the number of Discriminatory AI proposers across parameter sets. 
    Numerical calculations results are obtained using $M_P \in [0,100]$ with step 5, $h \in[0.4,0.6]$ with step 0.01, $l \in[0.1,0.3]$ with step 0.01, and $\beta \in \{0.1, 1, 10, 100\}$.}
    \label{figA4}
\end{figure}

\paragraph{Robustness across parameters.} Discriminatory AI proposers robustly sustain mutual fairness across a wide range of game payoffs and selection intensities, though with a reduced dominance of $HH$ compared to Samaritan AI. As shown in Figure \ref{figA3}a, the system still frequently converges to the mutually fair ($HH$) state, with the stationary distribution reaching 1.0 in approximately 50\% of the tested parameter combinations. While this represents a significant decrease from the roughly 88\% prevalence observed with Samaritan AI, the Discriminatory AI nevertheless prevents the system from being completely trapped in unfairness. Notably, the prevalence of the unfair $LL$ state (Figure \ref{figA3}d) remains limited, while the partially fair $HL$ and $LH$ states (Figure \ref{figA3}b, c) consistently fail to emerge as stable long-term outcomes, appearing only with a stationary distribution near zero. 

The prevalence of mutual fairness increases monotonically with the number of Discriminatory AI proposers. As illustrated in Figure \ref{figA4}, the stationary distribution of the fair $HH$ state exhibits an upward trend as the number of AI agents increases, and surpasses 0.9 with a relatively small agent pool. 

\begin{figure}
    \centering
        \includegraphics[width=1\linewidth]{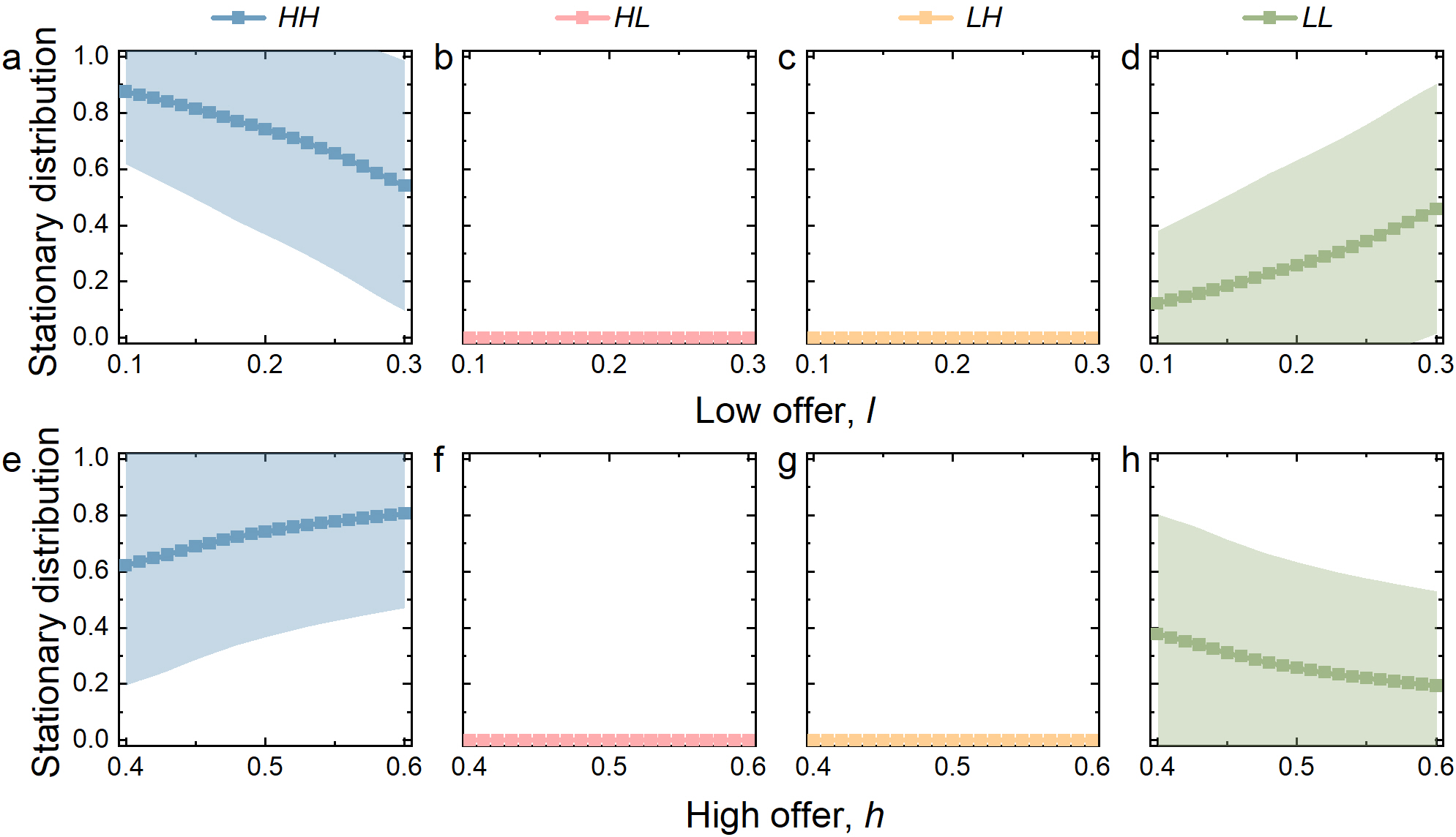}
    \caption{
    \textbf{Mutual fairness is robustly achieved when the low offer is minimal and the high offer is maximal.}
    Shown are the stationary distributions as a function of high offer and low offer across parameter sets. 
    Numerical calculations results are obtained using $M_P \in [0,100]$ with step 5, $h \in[0.4,0.6]$ with step 0.01, $l \in[0.1,0.3]$ with step 0.01, and $\beta \in \{0.1, 1, 10, 100\}$.}
    \label{figA5}
\end{figure}

Mutual fairness is most robustly achieved when the low offer is minimal, and the high offer is maximal, revealing the boundary conditions of Discriminatory AI effectiveness. As shown in Figure \ref{figA5}a and e, an increase in the low offer $l$ or a decrease in the high offer $h$ leads to a decline in the prevalence of the $HH$ state. 
Unlike Samaritan AI, which is less sensitive to payoff variations due to its unconditional nature, the success of Discriminatory AI is closely tied to the relative payoff advantage of fairness. Consequently, while Discriminatory AI is highly efficient, its performance is more dependent on the game's incentive structure, flourishing particularly when unfairness is easily punishable and fair offers remain attractive.

\section{Conclusion}
We developed an asymmetric Ultimatum Game model of hybrid populations to evaluate how role-specialised AI agents influence the evolution of human fairness. In our study, we first introduced the Samaritan AI agents, which  adopt hardcoded fair strategies as either proposers or receivers. By analysing the dynamics of finite well-mixed populations, we uncovered a fundamental asymmetry in the efficacy of these interventions: Samaritan receivers are significantly more potent at driving population-wide fairness than Samaritan proposers. Furthermore, we identified a synergistic effect: when Samaritan AI roles are introduced simultaneously, the critical threshold of AI agents required is substantially reduced. However, recognising the limitations of Samaritan proposers, especially in strong selection scenarios, we developed the Discriminatory AI proposer, which conditions its offers on receiver's predicted behaviour. This Discriminatory AI not only outperforms both Samaritan types in fostering global fairness but also drastically lowers the critical number of agents needed to sustain equity, particularly when selection pressure is strong. 

Our work reveals the decoupled evolutionary pressures of specialised AI, moving beyond the conventional assumption of role symmetry. Traditionally, research on the Ultimatum Game (UG) has relied on the assumption that individuals occupy both roles simultaneously \citep{rand2013evolution,nowak2000fairness,zisis2015generosity,santos2019evolution,cimpeanu2023social}; by contrast, our bipartite model introduces the crucial variable of role-specialisation. Our Samaritan agent demonstrates that AI's impact is role-dependent rather than a general cooperative catalyst. We show that placing AI in the norm-enforcing (receiver) position exploits the psychological mechanism of disadvantageous inequity aversion far more effectively than the modelling of fairness in the proposer position~\citep{fehr1999theory}. 
In addition, the superior performance of our Discriminatory AI in strong selection regimes suggests that fairness is most effectively promoted when interventions are targeted and leverage sanctioning-like mechanisms rather than uniform cooperation~\citep{cimpeanu2021cost}. By effectively sanctioning low-threshold receivers through reduced offers, Discriminatory AI creates a landscape rewarding fair behaviour. Our results highlight that AI's impact is role-dependent and strategy-specific, offering insightful guidance for AI design to maintain equity in resource-constrained hybrid environments.

Beyond the enhancement of fairness, the long-term deployment of AI in realistic interactions must account for the trade-off between institutional costs for AI development  and social welfare. Our results demonstrate that while both specialised AI agents are capable of sustaining fairness, their presence inherently introduces systemic costs that are often overlooked for the sake of model simplification. As maintaining behavioural patterns or implementing external interventions is rarely cost-free, these hidden expenditures can potentially diminish the overall social welfare even while successfully enhancing prosociality~\citep{han2025cooperation, song2025emergence}.
Therefore, a critical future challenge lies in balancing the AI-driven enhancement of fairness against the economic or social costs required to support such interventions.

Finally, our work establishes a foundation for several promising research directions.  Our study focused on well-mixed populations where individuals interact uniformly, while real-world social structures are often characterised by complex networks where interaction topology significantly influences norm diffusion \citep{szabo2007evolutionary,perc2013evolutionary}. Investigating the impact of specialised AI within our bipartite hybrid population model on structured populations, including scale-free, time-varying, and higher order networks, remains a vital next step. 
Additionally, while we employed Samaritan AI agents and Discriminatory AI agents to isolate the impact of role specialisation, future research should explore the effects of learning-driven agents, such as those utilising reinforcement learning or Large Language Models (LLMs). Such agents could dynamically adjust their thresholds or offers based on the human population's state, potentially uncovering more sophisticated, co-evolutionary pro-social behaviours in increasingly hybrid societies.
Furthermore, while our Discriminatory AI assumes perfect information regarding receiver strategies, real-world interactions are frequently plagued by informational noise and misjudgments. It's a promising direction to explore the impact of imperfect recognition (or strategic uncertainty) on the effectiveness of AI agents, for example by using intention recognition methods or LLMs \citep{sukthankar2014plan,wright2026assessing,han2013state}.

\ack{We acknowledge the support provided by EPSRC (grant EP/Y00857X/1) to Z.S. and T.A.H.}

\bibliographystyle{apalike}
\bibliography{mybib}

\end{document}